\documentstyle[12pt,amssymb,diagrams]{article}

\newcommand{\be}{\begin{eqnarray}}
\newcommand{\ee}{\end{eqnarray}}

\renewcommand{\d}{{\rm d}}
\newcommand{\dl}{{\delta}}

\normalbaselines 
\begin{document} 

\begin{center}
\vspace*{1.0cm}

{\LARGE{\bf Bicomplexes and finite Toda lattices}} 

\vskip 1.cm

{\large {\bf A. Dimakis${}^1$ and F. M\"uller-Hoissen${}^2$}} 

\vskip 0.5 cm 
${}^1$ Department of Mathematics, 
University of the Aegean \\
GR-83200 Karlovasi, Samos, Greece
\vskip.2cm \noindent
${}^2$ Max-Planck-Institut f\"ur Str\"omungsforschung \\
Bunsenstrasse 10, 
D-37073 G\"ottingen, Germany

\end{center}

\vspace{.1 cm}

\begin{abstract}
We associate bicomplexes with the finite Toda lattice and with a 
finite Toda field theory in such a way that conserved currents 
and charges are obtained by a simple iterative construction.
\end{abstract}


\noindent
In recent work \cite{DMH99a,DMH99b} we have demonstrated how bicomplexes can be associated with several completely integrable models in such a way that they provide us with an iterative construction of conserved currents. In the following we recall the underlying mathematical structure and demonstrate how the finite Toda chain and a finite Toda field equation fit into this scheme.
\vskip.2cm   

\noindent
Let $V=\bigoplus_{r \geq 0} V^r$ be an ${\mathbb{N}}_0$-graded linear space (over ${\mathbb{R}}$ or ${\mathbb{C}}$) and $\d , \delta \, : \, V^r \rightarrow V^{r+1}$ two linear maps satisfying
$\d^2 = 0$, $\delta^2 = 0$ and $\d \, \delta + \delta \, \d = 0$. 
Then $(V,\d,\delta)$ is called a {\em bicomplex}. In the following we assume that, for some $s \in {\mathbb{N}}$, $H^s_\delta (V)$ is trivial, so that all $\delta$-closed elements of $V^s$ are $\delta$-exact. Furthermore, we assume that there is a (nonvanishing) $\chi^{(0)} \in V^{s-1}$ with $\d J^{(0)} =0$ where $J^{(0)} = \delta \chi^{(0)}$.
Let us define $J^{(1)} = \d \chi^{(0)}$. Then $\delta J^{(1)} = - \d \delta \chi^{(0)} = 0$, 
so that $J^{(1)} = \delta \chi^{(1)}$ with some $\chi^{(1)} \in V^{s-1}$. Next we define
$ J^{(2)} = \d \chi^{(1)}$. Then $ \delta J^{(2)} = - \d \delta \chi^{(1)} = - \d J^{(1)} = - \d^2 \chi^{(0)} = 0$, so that $J^{(2)} = \delta \chi^{(2)}$ with some $\chi^{(2)} \in V^{s-1}$. 
This can be iterated further and leads to a (possibly infinite) chain (see Fig.~1) of $\delta$-closed elements $J^{(m)}$ of $V^s$ and $\chi^{(m)} \in V^{s-1}$ satisfying
\be
         J^{(m+1)} = \d \chi^{(m)} = \delta \chi^{(m+1)}   \; .
\ee
Introducing $\chi = \sum_{m \geq 0} \lambda^m \, \chi^{(m)}$ with a parameter $\lambda$, 
this becomes
\be
    \delta (\chi - \chi^{(0)}) = \lambda \, \d \, \chi \; .    \label{chi_eq}
\ee
In the following examples we will only consider the case where $s=1$. In these examples, the $J^{(m)}$ represent conserved currents which determine conserved charges. 

\small
\diagramstyle[PostScript=dvips]
\begin{diagram}[notextflow]
 &         &\chi^{(0)}&         &       &         &\chi^{(1)}&         &       &         &\chi^{(2)}&&& \\
 &\ldTo^\dl&          &\rdTo^\d &       &\ldTo^\dl&          &\rdTo^\d &       &\ldTo^\dl&          &\rdTo^\d&& \\
J^{(0)}&         &          &         &J^{(1)}&         &          &         &J^{(2)}&         &          & &J^{(3)}&\cdots\\
 &\rdTo^\d &          &\ldTo^\dl&       &\rdTo^\d &          &\ldTo^\dl&       &\rdTo^\d &          &\ldTo^\dl&&  \\
 &         & 0        &         &       &         & 0        &         &       &         & 0        &&&
\end{diagram}
\normalsize

\begin{center}
{\bf Fig. 1}  \\
The chain of $\delta$-closed $s$-forms $J^{(m)}$. 
\end{center}

\noindent
{\em Example 1:} The finite Toda lattice. \\
Let $C^\infty({\mathbb{R}},{\mathbb{R}}^n)$ be the set of smooth maps $f: {\mathbb{R}} \rightarrow {\mathbb{R}}^n$ and $\Lambda = \bigoplus_{r=0}^2 \Lambda^r$ the exterior algebra of a 2-dimensional vector space. We choose linearly independent 1-forms $\tau, \xi \in \Lambda^1$ which satisfy
$\tau^2 = \xi^2 = \tau \, \xi + \xi \, \tau = 0$. Furthermore, we set 
$V = \bigoplus_{r=0}^2 V^r = C^\infty({\mathbb{R}},{\mathbb{R}}^n) \otimes \Lambda$. 
Next we define linear maps $\d, \delta \, : \, V^0 \rightarrow V^1$ via
\be
    \d f = (L f) \, \tau + (\dot{f}+Mf) \, \xi  \, ,  \quad
    \delta f = \dot{f} \, \tau + (S f-f) \, \xi
\ee
with maps $S,M,L \, : \, {\mathbb{R}} \to M(n\times n; {\mathbb{R}})$ and 
$\dot{f} = d f/dt$ where $t$ denotes the canonical coordinate function on ${\mathbb{R}}$. 
$\d$ extends to $V^1$ via $\d (f \, \tau + g \, \xi) = (\d f) \, \tau + (\d g) \, \xi$ 
for all $f,g \in C^\infty({\mathbb{R}},{\mathbb{R}}^n)$, and correspondingly for $\delta$.
$\d$ and $\delta$ satisfy the bicomplex conditions iff
\be
    \dot{S} =0 \, , \quad \dot{M} = [S,L]  \, , \quad \dot{L} = [L,M] \; .
\ee
Let $e_i=(\delta^a_i)$ denote the standard basis of ${\mathbb{R}}^n$ and 
$E_{ij}=(\delta^a_i \delta_{jb})$ the elementary matrices. Then we have
$E_{ij} \, e_k = \delta_{jk} \, e_i$ and 
$E_{ij} \, E_{kl} = \delta_{jk} \, E_{il}$. Now we choose
\be
   S = \sum_{i=1}^{n-1} E_{i,i+1} \, , \quad
   M = \sum_{i=1}^{n} \dot{q}_i \, E_{ii} \, , \quad
   L = -\sum_{i=1}^{n-1} e^{q_i-q_{i+1}} \, E_{i+1,i} \; .
\ee
$S$ has the properties $\dot{S}=0$, $S e_1=0$, $Se_i=e_{i-1}$ for $i=2,\ldots,n$.
One finds that $\dot{L} = [L,M]$ is identically satisfied and
$\dot{M} = [S,L]$ is equivalent to the finite Toda lattice equation
\be
    \ddot{q}_1 = - e^{q_1-q_2} \, , \quad 
    \ddot{q}_n = e^{q_{n-1}-q_n} \, , \quad
    \ddot{q}_i = e^{q_{i-1}-q_i} - e^{q_i-q_{i+1}} \quad i=2,\ldots,n-1  \; .
         \label{finiteTodaEq}
\ee

$\delta J =0$ for $J = J_0 \, \tau + J_1 \, \xi$ means
$ S J_0-J_0 = \dot{J}_1 $. In particular, this implies $J = \delta \phi$ with 
$\phi = \sum_{k=1}^n \phi_k \, e_k$ and $\phi_k = -\sum_{i=k}^n J_{1i}$. 
Hence $\delta$-closed elements of $V^1$ are $\delta$-exact. 
Using the euclidean scalar product $\langle \, , \, \rangle$ and $u = \sum_{i=1}^n e_i$, we define 
\be
      Q = \langle u , J_1 \rangle
\ee
for a $\delta$-closed element $J \in V^1$. Then
\be
 \dot{Q} = \langle u,S J_0 -J_0 \rangle 
         = \langle S^t u-u , J_0 \rangle 
         = -\langle e_1 , J_0 \rangle  \; .
\ee
The $\delta$-closed elements $J^{(m)} \in V^1$ obtained via the above iteration 
procedure (for $s=1$) satisfy $J^{(m)} = \d \chi^{(m-1)}$ which implies 
$J^{(m)}_0 = L \chi^{(m-1)}$ and thus $\dot{Q}^{(m)} = 0$ since $L^t e_1=0$. 
Hence, the $Q^{(m)}$ are conserved.

Choosing $\chi^{(0)} = u$ (which satisfies $\d \delta \chi^{(0)} = 0$), the linear equation (\ref{chi_eq}) becomes equivalent to the
system
\be
  \dot{\chi} = \lambda \, L \chi \, , \quad
  (I-S) \chi = e_n - \lambda \, ( \lambda \, L \chi + M \chi )
\ee
where $I$ denotes the $n \times n$ unit matrix. Using 
$(I-S)^{-1} e_k = \sum_{j=1}^k e_j$, the last equation allows the recursive 
calculation of the $\chi^{(m)}$:
\be
 \chi^{(1)} &=& - (I-S)^{-1} M \chi^{(0)} = -\sum_{i=1}^n \sum_{j=1}^i \dot{q}_i \, e_j 
             = -\sum_{j=1}^n \sum_{i=j}^n \dot{q}_i \, e_j \\
 \chi^{(m)} &=& - (I-S)^{-1} ( M \chi^{(m-1)} + L \chi^{(m-2)} ) \quad m=2, \ldots,n \; .
\ee 
 From $J^{(m)} = \d \chi^{(m-1)}$ we obtain in particular
\be
  J^{(1)} &=& - \sum_{i=2}^n e^{q_{i-1}-q_i} e_i \, \tau 
              + \sum_{i=1}^n \dot{q}_i \, e_i \, \xi  \\
  J^{(2)} &=& - \sum_{k=1}^{n-1} e^{q_k-q_{k+1}} \, \chi^{(1)}_k
            \, e_{k+1} \, \tau + \sum_{k=1}^n ( \dot{\chi}^{(1)}_k 
            + \dot{q}_k \,   \chi^{(1)}_k) \, e_k \, \xi  \; .
\ee
The associated conserved charges are
\be
  Q^{(1)} = -\sum_{k=1}^n \dot{q}_k \, , \quad 
  Q^{(2)} = - {1 \over 2} \sum_{k=1}^n
    \dot{q}_k^2 - \sum_{k=1}^{n-1} e^{q_k-q_{k+1}} 
    - {1 \over 2} \left( Q^{(1)} \right)^2  \; .
\ee
To obtain the last expression, we made use of the equations 
of motion (\ref{finiteTodaEq}).
The conserved charges of the finite Toda lattices are well-known, of course \cite{Heno74}. On the $n$-point lattice, there are only $n$ independent conserved charges.
\hfill  \rule{7pt}{7pt}

\vskip.1cm
\noindent
{\em Example 2:} Finite Toda field theory. \\
Modifying the previous example, we now consider 
$V = C^\infty({\mathbb{R}}^2,{\mathbb{R}}^n) \otimes \Lambda$ 
and define linear maps $\d, \delta \, : \, V^r \rightarrow V^{r+1}$, $r=0,1$, via
\be
  \d f = (L f) \, \tau + (M f - f_x) \, \xi   \, , \quad
  \delta f = f_t \, \tau + (S f-f) \, \xi   
\ee
in terms of coordinates $t,x$ on ${\mathbb{R}}^2$. $f_t$ and $f_x$ denote the partial derivative with respect to $t$ and $x$, respectively. The bicomplex conditions are then equivalent to
\be
   S_t=0 \, , \quad M_t = [S,L] \, , \quad L_x= - [L,M] \; .
\ee
Choosing
\be
   S = \sum_{i=1}^{n-1} E_{i,i+1} \, , \quad 
   M = - \sum_{i=1}^n q_{i,x} \, E_{ii} \, , \quad
   L = - \sum_{i=1}^{n-1} e^{q_i-q_{i+1}} \, E_{i+1,i} \, ,
\ee
we obtain the field equations
\be
   q_{1,tx} \! &=& \! e^{q_1-q_2} \, , \quad q_{n,tx} = -e^{q_{n-1}-q_n} \, , 
                      \nonumber \\
   q_{i,tx} \! &=& \! e^{q_i-q_{i+1}}-e^{q_{i-1}-q_i}  \qquad i=2,\ldots,n-1 \; .
\ee
In the following we restrict our consideration to the case $n=2$ where this system 
is equivalent to the Liouville equation. Indeed, setting 
$q_1 = -q_2 = \phi$, we have
\be
 S = \left( \begin{array}{cc}0&1\\0&0\end{array} \right) \, , \quad
 M = \phi_x \left(\begin{array}{cc}-1&0\\0&1\end{array} \right)  \, , \quad
 L = -e^{2 \phi} \left( \begin{array}{cc}0&0\\1&0\end{array} \right)  \, ,
\ee
and the field equations reduce to 
\be
     \phi_{tx} = e^{2 \phi}  \; .
\ee
Writing $\chi = (a , b )^t$, the linear system (\ref{chi_eq}) with 
$\chi^{(0)} = u$ yields
\be
   a_t = 0 \, , \qquad
   b_t = - \lambda \, a \, e^{2 \phi}
\ee
and
\be
   b = 1 + \lambda \, (b_x - \phi_x \, b)  \, , \qquad
   b = a - \lambda \, (a_x + \phi_x \, a)  \; .   
                \label{ab-rels}
\ee
As a consequence, we have
\be
   a - 1 - 2 \lambda \, a_x + \lambda^2 ( a_{xx} + v \, a ) = 0
\ee
where we introduced 
\be
    v = \phi_{xx}-\phi_x{}^2   \label{vphi}
\ee
which satisfies $v_t = 0$ as a consequence of the Liouville 
equation.\footnote{Actually $v_t=0$ is equivalent to the Liouville 
equation as long as $\phi_{tx} \neq 0$. Regarding (\ref{vphi}) as an 
equation for $w = \phi_x$ with a given function $v(x)$,
we recover the Ricatti equation $ w_x - w^2 - v(x) = 0 $
associated with the Liouville equation.} 
The conserved current is
\be
  J = \d \chi 
    = - e^{2 \phi} \, a \, \left( \begin{array}{c}0\\1\end{array} \right) \, \tau 
    - \left( \begin{array}{c} a_x + \phi_x \, a \\ b_x - \phi_x \, b
      \end{array} \right) \, \xi
\ee
and the associated conserved charge (constructed as in the previous example) becomes
\be
  Q = - (a_x + \phi_x \, a) - (b_x - \phi_x \, b) = \lambda^{-1} \, (1-a)   \label{Q-a}  
\ee
with the help of (\ref{ab-rels}). Note that $a_t=0$ implies directly $Q_t=0$. 
The corresponding conserved charges $Q^{(m)}$ are
\be
  Q^{(1)}=0 \, , \quad Q^{(2)}= v \, , \quad Q^{(3)}= 2 \, v_x \, , \quad
  Q^{(4)}= 3 \, v_{xx}-v^2
\ee
and so forth. From the above formula for $a$ it is obvious that all the $Q^{(m)}$ 
depend on $\phi$ only through $v$. Since $v$ is conserved, the $Q^{(m)}$ are trivially conserved in this example.

Above we used a special solution of $\d \delta \chi^{(0)} = 0$. Its general solution is
\be
   a^{(0)} &=& c_1(x) + \int^t c_2(t') \, e^{-\phi} \, dt'  \\
   b^{(0)} &=& \left( c_4(t) + \int^x [ \, c_3(x') - \int^t a^{(0)} \, e^{2 \phi} \, d t' \, ] 
               \, e^{-\phi} \, dx' \right) \, e^\phi 
\ee
with arbitrary differentiable functions $c_1, \ldots, c_4$ of a single variable. 
(\ref{Q-a}) has to be replaced by $Q = \lambda^{-1} (a^{(0)} - a)$. Our construction 
of conserved quantities does not really lead to anything new, however. In particular, 
we get $Q^{(1)} = h(x)$ with an arbitrary function of $x$ (related to $c_3$) and 
$Q^{(2)} = c_{1,xx} + 2 \, h_x + c_1 \, v$. The corresponding recursion formula is
$Q^{(m)} = 2 \, Q^{(m-1)}_x - Q^{(m-2)}_{xx} - v \, Q^{(m-2)}$ for $m >1$.
\hfill  \rule{7pt}{7pt}
\vskip.2cm

We expect that convenient bicomplexes can also be associated with generalizations (see \cite{Chao95}, for example, and the references therein) of the above Toda systems in order to generate their conserved currents and charges.

\section*{Acknowledgment} 
 F. M.-H. profitted quite a lot from participation in a long series of 
conferences on modern topics in mathematical physics organized by 
Professor H.-D. Doebner and his collaborators over the years. On the occasion of Professor Doebner's retirement, it is a pleasure to 
thank him for his support and his outstanding promotion of 
mathematical physics especially in Germany.

\end{document}